\documentclass[twocolumn,tightenlines,showpacs,aps]{revtex4}
\usepackage{graphicx}
\usepackage{t1enc}
\usepackage[latin1]{inputenc}
\usepackage[english]{babel}

\begin{document}

\title{Vaccination pattern affects immunological response}

\author{P. G. Etchegoin}
\email{Pablo.Etchegoin@vuw.ac.nz}

\affiliation{The McDiarmid Institute for Advanced Materials and Nanotechnology\\
School of Chemical and Physical Sciences\\ Victoria University of Wellington\\
PO Box 600 Wellington, New Zealand}

\date{\today}

\begin{abstract}
The response of the immune system to different vaccination
patterns is studied with a simple model. It is argued that the
history and characteristics of the pattern defines very different
secondary immune responses in the case of infection. The memory
function of the immune response can be set to work in very
different modes depending on the pattern followed during
immunizations. It is argued that the history and pattern of
immunizations can be a decisive (and experimentally accessible)
factor to tailor the effectiveness of a specific vaccine.
\end{abstract}


\maketitle

\section{Introduction and overview}

From the physical scientist standpoint, the immune system
(IS)\cite{Wise,Eales} (with cell-mediated and/or humoral
responses) ranks amongst the most complex naturally-occurring
nonlinear many-body problems we can find\cite{Perelson}. The IS
involves interactions amongst several entities (antibodies (ATB),
antigens (ATG), immune complexes (IC), natural-killer (NK) cells,
plasma cells (PC), T-cells, B-cells, antigen presenting cells
(APC), etc) with complex affinities and dynamics. The
understanding of the IS response, even at a qualitative level, is
of prime importance for current issues in immunology, ranging from
HIV\cite{Nowak} and other immunodeficiencies, to tumor
immunotherapy\cite{Rao}. Some of the components of the IS (like T,
NK, or B-cells) have complex internal dynamics of their own,
resulting in differentiation, division, and mutation. The dynamics
of the IS presents a problem with a hierarchy of complexities at
different levels; the IS response is to a cell what sociology is
to an individual, having both internal and emergent properties
arising from their inherent complexity.

Mathematical modelling in theoretical immunology has come as an
aid in the understanding of the IS. This is in part due to the
unprecedented growth in computer memory and processor speed, but
also due to a better understanding of the dynamics of the IS.
Extensive reviews of the early pioneering work in the mathematical
modelling of the IS can be found in the literature\cite{Perelson}.

Admittedly, in spite of more than two decades of research, the
modelling of the IS response is still in its infancy. The shear
complexity of the problem is not the only reason, but also the
fact that many microscopic interactions among different components
are either not fully understood, difficult to measure, or no
reasonable parametrization is known for them. There is also the
widespread opinion that not all of the fundamental molecules
participating in the IS response might have been fully
identified\cite{Perelson}, in particular compounds related to
inter-cell signalling and communication.

Despite all these shortcomings, the development of models is an
important advance in our understanding and a clear aid in the
development of intuition and strategies to guide the IS in the
right direction to combat diseases like AIDS or cancer. Monoclonal
antibodies (MA) and interferon based therapies have become
nowadays part of the standard repertoire in cancer treatment,
while many other types of immunotherapy, such as cancer vaccines
(CV), remain largely experimental. CV's consist in most cases of a
source of cancer-related material (antigen) which is injected to
further stimulate the IS. The experimental challenge so far has
been to find better antigens with enough effectiveness to enhance
the patient's IS to fight cancer cells.

This paper focuses on a very specific aspect of the IS: the
amplitude of the secondary immune response according to different
{\it training} programs established in the vaccination pattern
(primary response). In general terms, the most basic task of the
immune system is pattern recognition; a task which the IS achieves
through the mechanism of {\it clonal selection}, elucidated more
than 40 years ago\cite{MacFarlane}. Clonal selection prepares the
{\it memory} of the IS to deliver a strong response (the secondary
response) if the antigen reappears. The initial steps of the
secondary response are crucial to the faith of the organism in its
ability to fight a recurring antigen. If the IS has not been
trained before, a massive attack from an antigen leaves the
organism to rely only on the primary response. This has the
disadvantage that there is an associated delay, because the cell
population needs to enlarge before substantial amounts of
antibodies can be produced. This is partly the reason why
illnesses from highly mutating antigens like influenza are more
difficult to fight by the IS than more severe, but at the same
time more stable, antigens. The former leaves the IS at the mercy
of the primary response, while the latter can be more effectively
fought through vaccination.

The way the IS develops a {\it memory} is based on its training
(vaccination) program. It will be argued that the ability to
develop a memory, like in many other pattern recognition problems,
is strongly dependent on the type of training. The possibility of
playing with the timing and sequence of the vaccination program to
tailor and maximize the effectiveness of the IS secondary response
is suggested.

\section{The immune system as a learning machine}

On very general theoretical grounds the IS is a learning machine
for pattern recognition of the epitopes of antigens. Forrest {\it
et. al}\cite{Forrest} have considered the adaptability of the IS
response as a pattern recognition and learning process using a
genetic algorithm\cite{Rojas} on a binary string model. After Ref.
\cite{Rojas}, genetic algorithms are stochastic search methods
managing a population of simultaneous search positions; they
evaluate the target function to be optimized at some randomly
selected points of the definition domain. At this level of
abstraction, there is a strong overlap between the learning
properties of the IS and many concepts in stochastic neural
networks, including learning algorithms and optimization. As far
as the experimental evidence is concerned, there is no clear-cut
demonstration on the exact algorithm the IS uses to learn. Other
options different from genetic algorithms like Boltzmann or
Hebbian learning\cite{Rojas} should be considered on an equal
foot.

Several radically different types of models have been proposed for
the IS, going from coupled systems of (nonlinear) differential
equations\cite{Perelson,1,2}, to spin-glasses\cite{Parisi1}, to
cellular automata (CA)\cite{Seiden1,Seiden2}. Segel pointed
out\cite{Segel} that a hallmark of complex phenomena is that they
can not be modelled by a single approach or, alternatively, that
they are prone to several different representations, depending on
the specific aspect we want to understand. The aspects of
training, learning, and pattern recognition we want to study are
more easily implemented on CA versions of the IS. We shall use,
accordingly, one of the well established CA models of the immune
response\cite{Seiden1,Seiden2} to address the issues we raised
before, trying to draw analogies with aspects of learning in
neural networks where appropriate.

For the sake of argument, consider the case of Boltzmann learning
in a simulated annealing process of the type used for neural
networks\cite{Rojas}. Simulated annealing, which is a special case
of the Monte Carlo method, was brought into the mainstream of
numerical optimization of networks after the seminal paper by
Kirkpatrick and coworkers\cite{Kirkpatrick}. The learning process
has here a direct physical meaning: it is equivalent to the
previous thermodynamic history of the network in its search for
the global minimum of the total energy. This is the clearest
example in which it is obvious that the outcome of the learning
process strongly depends on the thermodynamic cycles and patterns
followed in the annealing. One quick temperature jump followed by
a rapid quench will have a completely different effect than a
sustained and gradual temperature drop, even if the amount of
energy put in is the same. The differences will be more pronounced
the more complex the energy landscape or the network. It is
objective of this paper to show that the same phenomenon exist in
simple models of the IS, thus showing the importance of the
pattern followed in the training period (vaccination).

\section{The model}
We do not to validate a new model for the IS here. Instead we
adopt a well established CA-model for the IS (IMMSIM), first
introduced by Celada and Seiden\cite{Seiden1} and further
developed by Kleinstein and Seiden\cite{immsim}. We comment very
briefly on the model and refer to the published
literature\cite{Seiden1,Seiden2,immsim} for the details.
Essentially, the model follows the evolution in a CA (spatial
lymph-node) of: antigen presenting cells, antibodies, antigens,
B-cells, immune complexes (IC), plasma cells, and T-cells. We use
the 8-bit string implementation with the parametrization suggested
by Kleinstein and Seiden\cite{immsim}. It is well known that even
the simplest automata with the simplest interaction rules can lead
to extremely complex behavior including chaotic dynamics, unstable
periods, and complex spatiotemporal patterns\cite{Wolfram}. An
automata like the one used in IMMSIM has the additional
complication that some of its entities change over time in a
stochastic manner, from hyper-mutations to the finite lifetime of
the cells; its dynamics can only be assessed by direct simulation
on a grid. Further details of the model can be found in the
original papers\cite{Seiden1,Seiden2,immsim}. It is assumed that
all memory cells (APC, B-, and T-cells) have an infinite lifetime
while the non-memory version of the same cells have an average
half-life of $\tau=$10 cycles, defined as the probability $P={\rm
exp}(-{\rm ln}2/\tau)$ to die in each cycle. The
permanence/extintion of a cell is decided stochastically in each
iteration, a feature that transforms the IMMSIM model into an
stochastic cellular automata. Direct interaction between IC's and
B-cells are not considered in the
model\cite{Seiden1,Seiden2,immsim}. The relative shorter lifetimes
of B-cells as compared to the total vaccination period makes the
total number of B-cells to be dominated mostly by memory B-cells.
We shall use, accordingly, the total number of B-cells as a
measure of the gained immune memory in the next section.

\section{Results and discussion}

We adopt a few definitions: We define two periods $(a)$ a
vaccination/inoculation or training period in which the CA is
trained by exposing it to a small quantity of antigen with
different temporal patterns, and $(b)$ an infection period in
which the CA is exposed to much larger quantities of antigen which
is injected at regular short intervals in the dynamics; thus
simulating a quasi-continuous production of antigen caused by
illness or external agents.

We will judge the ability of the IS to overcome infection by its
ability to reduce the amount of existing antigen to exactly zero,
during the infection period. This is a somewhat arbitrary
definition but it will reveal, precisely, the different types of
behavior that can be generated in the CA according to the training
pattern.

Form the experimental point of view, once an antigen or a modified
form of antigen has been produced for a trial, there are very few
variables left except for: $(a)$ the amount of antigen to be
introduced as vaccine, and $(b)$ the number of inoculations and
the length of the training program. The results in this paper
suggest that for exactly the same amount of antigen used as a
vaccine, the IS may respond in completely different ways depending
on how that vaccine is spread over the training period.

\begin{figure}
\begin{center}
\includegraphics[width=8cm]{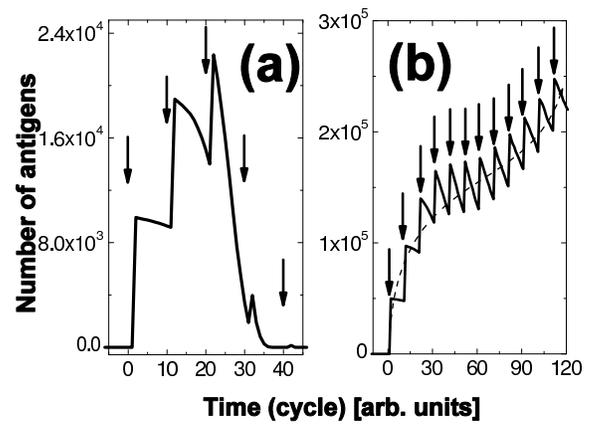}
\caption{(a) Primary response of an outbreak of 10$^4$ antigens
every 10 cycles in a non-vaccinated CA. The system copes with the
outbreak and reduces the amount of circulating antigen to zero
after$\sim 35$ cycles. In (b) the outbreak is of $5\times10^4$
antigens every 10 cycles. There is a small slowdown in the
increase of the antigen population after the initial response of
the IS, but the infection eventually wins and the antigen
population increases indefinitely. This would be a case of an
infection overcoming the capacity of the IS to cope with an
unknown antigen only by means of its primary response.}
\label{fig1}
\end{center}
\end{figure}

We first concentrate on the concepts of infection and primary
response for the purpose of the modelling here. Figure \ref{fig1}
shows the example of of a CA which {\it has not} been vaccinated,
i.e. it has no previous recollection of an encounter with the
antigen, and it is exposed at some point to an infection in the
terms defined above. The CA has a standard population of T-, APC,
and B-cells of$\sim 10^3$ and no antigen is present initially.
Here we are testing the ability of the model-IS to cope with an
infection by means of its primary response only. In a CA the
cycles play the role of time; they can be used as synonyms in this
context. In Fig. \ref{fig1}(a) the CA is injected from $t=0$
onwards with 10$^4$ antigens every 10 cycles. The positions where
the antigens are introduced are marked with vertical arrows. In
Fig. \ref{fig1}(a) we see that after 4 exposures to the periodic
outbreak of antigen the CA has managed to reduce the antigen
content to zero. This is a case where, with some delay, the IS is
able to control the infection by means of its primary response. On
the contrary Fig. \ref{fig1}(b) shows an example where the primary
response cannot cope with the outbreak. Here we inject 5$\times$
10$^4$ antigens every 10 cycles and monitor again the amount of
antigen present in the CA. After a rapid increase, there is a hint
of a slowdown in the amount of antigen present, i.e. the primary
response is responding and fighting back partially the infection.
But eventually the constant increase of antigen at regular times
wins and the amount of antigen increases continuously and cannot
be controlled. We are interested now on how a vaccination program
with a small amount of antigen can help the IS to cope with this
latter situation.

\begin{figure}
\begin{center}
\includegraphics[width=8cm]{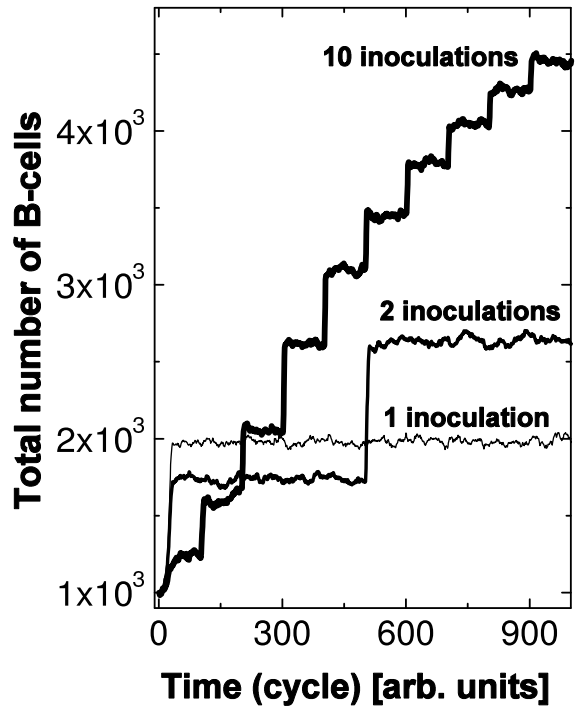}
\caption{Total number of B-cells as a function of time (cycle) for
a $10^3$-cycles vaccination program with $10^4$ antigens. Three
cases are shown: 1 inoculation with the full dose of 10$^4$ at
$t=0$, 2 inoculations at $t=0$ and 500 with $5\times 10^3$
antigens, and 10 inoculations with $10^3$ antigens every 100
cycles. The total population of B-cells is dominated by memory
B-cells which have an infinite lifetime (unlike normal cells with
a half-life of 10 cycles). The distributed vaccination program
achieves a total memory for a secondary response which is more
than twice the memory gained with a single inoculation.}
\label{fig2}
\end{center}
\end{figure}

We now define a vaccination program. From the response in Fig.
\ref{fig1} we can deduce that 10$^3$ iterations is a very long
time compared to the response of the CA. In addition, an injection
of a total of 10$^4$ antigens spread over $10^3$ cycles can be
easily handled by the model-IS. We define a vaccination program,
accordingly, based on a total antigen intake of $10^4$ (the
vaccine) spread in different ways over a period of $10^3$ cycles
(the training period). Figure \ref{fig2} shows the result of
different vaccination patterns. A measure of the ability of the
system to memorize the presence of the antigen for future
infections is the total number of B-cells present in the system.
As one of the primary targeting centers of antigens, B-cells play
a decisive role on the memory of the IS to trigger a fast
secondary response. The total number of B-cells is dominated by
the memory B-cells, as explained before.

Figure \ref{fig2} shows then the B-cell population resulting from
3 different vaccination patterns with the same total amount of
antigen: $(a)$ a single inoculation with 10$^4$ antigens at $t=0$,
$(b)$ two inoculations with $5\times 10^3$ antigens at $t=0$ and
$t=500$ cycles, and $(c)$ 10 inoculations with $10^3$ antigens
every 100 cycles. It is evident from the B-cell population that
the build up of immune memory is very different depending on the
pattern followed during vaccination, even if the total amount of
antigen is the same in all cases. Smaller doses spread-out over
the training period result in better memory retention as far as
the B-cell population is concerned; but the picture of the better
memory gain with small doses is in fact subtler. A naive
conclusion from Fig. \ref{fig2} would be that we have to continue
making the doses smaller and more frequent. But there is in fact a
tradeoff between the amount, frequency, and lifetime of the
antigens. The gain in B-cell population for 20 inoculations with
half of the dose is negligible, for example, and for 50
inoculations the final population of B-cells after vaccination is
smaller than in the 10 inoculation program, as we shall show
later.

\begin{figure}
\begin{center}
\includegraphics[width=8cm]{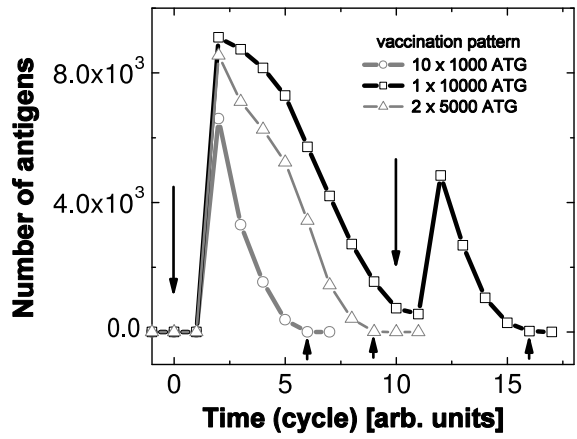}
\caption{Response of the IS to the infection in Fig. \ref{fig1}(a)
after vaccination. For the single inoculation case the infection
is controlled only after the second antigen injection around$\sim
16$ cycles. The small vertical arrows at the bottom of the curves
show the places where the infection is controlled for the first
time in the three cases. The two and 10 inoculations program
control the infection within the first injection, but with marked
differences in response time. See the text for further details.}
\label{fig3}
\end{center}
\end{figure}

The three different scenarios in immune memory gain in Fig.
\ref{fig2} result naturally in three completely different
secondary immune responses in the event of infection. Figure
\ref{fig3} exemplifies this with an infection of the type shown if
Fig. \ref{fig1}(a), i.e. 10$^4$ antigens every 10 cycles. The IS
controls the infection now much faster than the$\sim 35$ cycles it
takes to the primary response, as expected, but significant
differences are seen in the dynamics of the three cases. The
single inoculation training controls the infection only after the
second antigen injection, while the 2 and 10 inoculations
approaches, control the outbreak right after the first appearance
of the antigen and before the second injection. During that time
the dynamics manages to clear the CA from antigen; the 10
vaccination approach being the fastest. It may seem {\it a priori}
obvious that the more vaccines the better the protection, but let
us stress that all vaccination programs are carried out with
exactly the same amount of antigen; i.e. more inoculations does
not mean more vaccine, but rather the same amount of vaccine
distributed differently.

\begin{figure}
\begin{center}
\includegraphics[width=8cm]{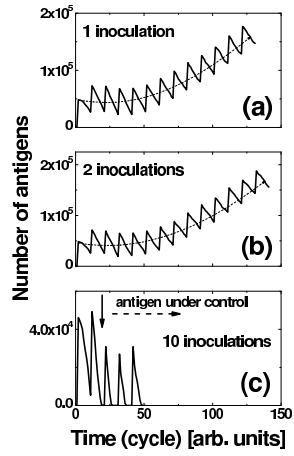}
\caption{Same as Fig. \ref{fig3} but for the infection in Fig.
\ref{fig1}(b). The three cases of vaccination are shown in
different plots for clarity. This is an infection which cannot be
controlled by the primary response, as shown in Fig.
\ref{fig1}(b). The one and two inoculations programs ((a) and (b))
cannot control the outbreak after vaccination despite a small
slowdown in the antigen population at shorter times. The infection
prevails in the long run and the antigen quantity increases
indefinitely. The 10-dose program (see Fig. \ref{fig2}) in (c)
succeeds to control the antigen population after two injections.
Further injections of antigen are controlled before the next one
comes in; three successive injections after the second are shown
in (c). The vertical arrow shows the place where the infection is
controlled for the first time. The IS is in complete control of
the outbreak in this case. For the {\it same} amount of vaccine,
the difference between life and death for this model IS resides in
the way the vaccine is implemented.} \label{fig4}
\end{center}
\end{figure}

The effect of the different vaccination programs and the different
immune memory gains is more dramatic in outbreaks that cannot be
controlled by the primary response of the IS. This is explicitly
shown in Fig. \ref{fig4}: a CA trained with the vaccination
programs of Fig. \ref{fig2} is exposed now to an infection
outbreak of $5\times10^4$ antigens every 10 cycles. Figures
\ref{fig4}(a) and (b) show the responses of the single and double
vaccination programs. After an initial increase, there a
sub-linear slowdown in the population of antigen which is a
mixture of primary and secondary response. But the outbreak
eventually dominates and the antigen population increases out of
control from there on. The 10 inoculations program, on the other
hand, controls the outbreak immediately after the second antigen
injection. We show a few more cycles after the second to
demonstrate that the outbreak effectively remains under control
and that the model-IS manages to reduce the antigen population to
zero in between injections. The difference in the training program
during vaccination represents, in this model, the difference
between life and death for the IS.

\begin{figure}
\begin{center}
\includegraphics[width=8cm]{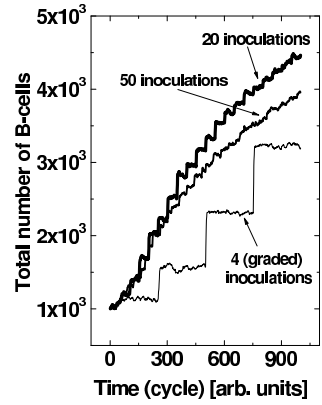}
\caption{Effect of further dilution in the vaccination doses. A 20
vaccine program attains approximately the same number of B-cells
than a program based on 10 vaccines (compare with Fig.
\ref{fig2}). A 50-vaccinations programm with the same total number
of antigens achieves a total memory on B-cells which is even
smaller. The figure shows also a program with four doses of 500,
1500, 2500, and 5500 antigens at $t=0$, 250, 500, and 750 cycles,
respectively. The unsurpassed result of the 10 vaccinations
program in Fig. \ref{fig2} is achieved by a compromise among:
small primary response per dose, lifetime of non-memory B- and
T-cells and APC's and minimum secondary response from the
accumulated memory. See the text for further details.}
\label{fig5}
\end{center}
\end{figure}

The important question is then how to maximize the memory of the
IS, given a certain amount of vaccine. As pointed out before, it
is not as simple as spreading out the dose as much a possible over
the training period by means of more frequents inoculations with
less amount of antigen. Let us first show explicitly the behavior
of the CA for more frequent vaccinations. Figure \ref{fig5} shows
the increase in B-cell population for a 20 and 50 inoculations
program with the same total amount of vaccine used in Fig.
\ref{fig2}. It can be appreciated in the figure that a 20
inoculations program finishes with the same number of B-cells as
the training with 10 vaccines (compare with Fig.\ref{fig2}) and,
moreover, a 50 inoculations treatment results in a smaller number
of B-cells than the previous ones. An obvious alternative is a
training program with a graded vaccination dose of antigen. Figure
\ref{fig5} shows an example where the $10^4$ antigens used as
vaccine are spread in four doses of 500, 1500, 2500, and 5500 at
$t=0$, 250, 500, and 750 cycles, respectively. This an other
similar vaccination patterns cannot surpass the total immune
memory gain of the 10 vaccine training.

From here we are left with the question of: why does the immune
memory increases by spreading the vaccination antigen evenly
during the training period, up to a maximum number of doses above
which the effect reverses? The answer to this comes from a
combination of effects: the lifetime of the cells, the primary
response, and the cumulative immune memory gained during the
vaccination period. The ideal vaccination program should aim at
having: $(i)$ a minimum of primary response per vaccination,
$(ii)$ a minimum effect of secondary response from the previous
vaccines, and $(iii))$ a maximum memory imprint on future
immunological responses. The gain of memory, in particular,
becomes more and more difficult when a previous history of
vaccination is already present, for the new dose is directly
exposed to a secondary response with production of antibodies and
rapid elimination of the new antigen. This explains the slowdown
in the immune memory gain seen in Fig. \ref{fig5}. For real immune
systems, the tailoring of the dose has to be decided
experimentally. In particular, it depends strongly on the type of
antigen, which is some cases is difficult to obtain (like in
cancer vaccines) and may not provoke a full immunological
response. A fine balance between the effectiveness of different
vaccination programs and the type of antigen can only be achieved
through experimental studies on large populations. Real systems
have also outbreaks with {\it continuous} production of antigen
rather than the "model-outbreak" studied here. We observed
qualitatively similar results by spreading the antigen in the
outbreaks in different manners, but it is not obvious in general
that a more complex or realistic automata will not display
different dynamics depending on the nature of the outbreak.

The important qualitative lesson learnt from the simulations here,
however, is that the timing of the vaccination program may provide
an additional variable to increase the effectiveness of vaccines.
The effect exist, at least, in a basic model of the IS with the
essential variables and components.

\section{Conclusions}
The effect of different programs of exposure to antigen in a model
IS have been investigated with the aim to understand the
mechanisms to achieve maximum immune memory for a fixed amount of
antigen. The gain of immune memory is a delicate compromise
between the increase in the population of defending cells and the
generation of antibodies through the fast secondary response which
tends to block the effect of further doses. As far as the
prediction of simple cellular automata is concerned, the results
in this paper demonstrate that multiple small doses of vaccine
with the same total amount of antigen can boost the immune memory
to the extent that it can make the difference between survival or
death for the organism. A separate study of long-term effects of
the vaccination program on the immune response is in progress and
will be published elswhere\cite{yo}.

\newpage

\end{document}